\documentclass[12pt]{JHEP3}
\usepackage{amsmath,amssymb,bbm}
\usepackage{multirow,colordvi}

\def\2{\frac12}
\def\4{\frac14}

\def\g{\gamma}

\def\s{\sigma}

\def\t{\tau}

\def\c{\chi}

\def\be{\begin{equation}}
\def\ee{\end{equation}}
\def\bea{\begin{eqnarray}}
\def\eea{\end{eqnarray}}
\author{Eric A. Bergshoeff, Mees de Roo, Sven F. Kerstan\\Centre for
Theoretical Physics, University of
Groningen, Nijenborgh 4, 9747 AG Groningen, The Netherlands\\
\email{E.A.Bergshoeff, M.de.Roo, S.Kerstan@phys.rug.nl}}
\author{Tom\'as Ort\'\i n\\Instituto de F\'\i sica Te\'orica
UAM/CSIC, Facultad de Ciencias C-XVI, C.U. Cantoblanco,
E-28049-Madrid, Spain\\ \email{Tomas.Ortin@cern.ch} }
\author{Fabio Riccioni
\\ DAMTP, Centre for Mathematical Sciences,
University of Cambridge,  Wilberforce Road, Cambridge CB3 0WA, UK\\
and\\
 Dipartimento di Fisica, Universit{\`a} di Roma ``Tor
Vergata'', I.N.F.N. - Sezione di Roma II ``Tor Vergata'', Via della
Ricerca Scientifica, 1 - 00133 Roma - ITALY\\
\email{F.Riccioni@damtp.cam.ac.uk}}

\abstract{We calculate the tensions of all
half-supersymmetric
nine-branes in IIB string theory. In particular, we point out the
existence of a solitonic IIB nine-brane. We
find that the D9-brane and its duality transformations parametrize
a two-dimensional surface in a four-dimensional space.}

\preprint{UG-06-01\\IFT-UAM/CSIC-06-02\\ROM2F/06-02} 

\title{IIB Nine-branes}

\begin{document}
\section{Introduction}

Using the supersymmetry algebra, there is a standard procedure to
construct the supergravity multiplets of IIA and IIB supersymmetry.
Naturally, the field content of these multiplets is such that there
is an equal number of (on-shell) bosonic and fermionic degrees of freedom.
However, it turns out that additional bosonic spacetime fields, which do
not describe propagating degrees of freedom, can be added
to these multiplets. They play an important role in describing the coupling of
supergravity to branes.

An example of this phenomenon is the addition of a nine-form
potential to the IIA supergravity multiplet \cite{Bergshoeff:1996ui}
which couples to a D8-brane. Integrating out this potential leads to
an integration constant that can be identified as the cosmological
constant parameter of massive IIA supergravity \cite{Romans:1985tz}.
Another example is the addition of a RR ten-form potential to the
IIB supergravity multiplet that couples to a D9-brane
\cite{Polchinski:1995mt}.  From a string theory point of view,
D9-branes take part in obtaining ten-dimensional type-I string
theory from type-IIB. Indeed, in the closed sector the orientifold
projection~\cite{augusto} generates a tadpole, and tadpole
cancellation, {\it i.e.} cancellation of the overall RR charge and
tension, requires the introduction of an open sector, corresponding
to D9-branes. The ten-form potential corresponding to these branes
does not have any field strength, and therefore there is no
supergravity solution corresponding to the D9-brane. Nevertheless,
the RR ten-form potential can consistently be included in the
supersymmetry algebra, and its gauge and supersymmetry
transformations formally determine the world volume action of the
D9-brane, from which the open sector of the low-energy action of the
type-I theory is obtained after truncation. In
\cite{Bergshoeff:1999bx} it was shown that the IIB algebra can be
extended in order to include the RR ten-form and a second ten-form
potential. Somewhat surprisingly, it has been pointed out that these
potentials cannot be related by an $SL(2,\mathbb{R})$ transformation
\cite{fabio}.

Recently, these issues were sorted out when it was shown that the ten-form that
couples to the
D9-brane transforms as a component of a quadruplet representation
under the duality group $SL(2,\mathbb{R})$ \cite{Bergshoeff:2005ac}
whereas the second ten-form potential of \cite{Bergshoeff:1999bx}
transforms as the component of a separate doublet representation.

Each of these ten-form potentials couples to a nine-brane.
It is the purpose of this letter to calculate which of these
nine-branes correspond to half-supersymmetric BPS objects and to
calculate their  tensions.

\section{Brane Tensions and BPS conditions}

To calculate the tension of a single p-brane and to determine the BPS condition
it is convenient to consider the leading terms of the full kappa-symmetric
brane action, assuming that such an action exists.
To be precise we consider (in static gauge) the Nambu-Goto term and the
term that describes the coupling of a $(p+1)$-form potential
$A_{(p+1)}$ to
the p-brane:
\begin{equation}\label{braneL}
{\cal{L}}_{\rm brane}\, =\, \tau_{\rm brane}\, \sqrt{-g} +
\epsilon^{\mu_1\cdots \mu_{p+1}}\,
A_{\mu_1\cdots \mu_{p+1}}\,.
\end{equation}
Here $\tau_{\rm brane}$ is the brane tension which in
general is a function of the scalars at hand (a dilaton
for IIA and a dilaton plus axion for IIB).

We now consider the supersymmetry variation of the brane action
\eqref{braneL}, but keep only terms linear in the gravitino\footnote{
A similar variation was considered in
\cite{Bergshoeff:2000zn} in the context of a supersymmetric
Randall-Sundrum scenario, and was also discussed in \cite{Bandos:2001jx}.}.
In this letter we restrict ourselves to the
IIB theory. We will consider the IIA case in an upcoming work
\cite{upcoming}. The relevant
supersymmetry variations of the spacetime metric $g_{\mu\nu}$ and
the $(p+1)$-form $A_{(p+1)}$ are given by\footnote{
We work with real spinors in string frame, see section 6 of
\cite{Bergshoeff:2005ac}. All spinors are two-component vectors,
with each component being a 16-component Majorana-Weyl spinor.}
\begin{equation}
\delta\, g_{\mu\nu}\, =\, 2i\bar\epsilon \gamma_{(\mu}\psi_{\nu)}
+ {\rm h.c.}\, ,
\hskip 1truecm
\delta A_{\mu_1\cdots \mu_{p+1}} \, \sim\, f\,\bar\epsilon
\gamma_{[\mu_1\cdots \mu_p} \sigma\psi_{\mu_{p+1}]}\,,
\end{equation}
where $f$ is a function of the dilaton plus
axion and $\sigma$ is one of the three Pauli-matrices\footnote{This is
not the generic situation. Sometimes
we are dealing, in the same supersymmetry variation,
with two terms each containing a different Pauli matrix and
a different dependence on the scalars.
 In that case the formulae below need a slight modification.
As an example of such a situation, see the discussion of the D1-brane
below.}.
For a half-supersymmetric
p-brane we require that the brane action is invariant under 16
of the 32 linear IIB supersymmetries (in addition there will be
16 nonlinear supersymmetries). For this to happen it is mandatory that the
supersymmetry variation of the brane action is proportional to a
projection operator  that projects
away half of the supersymmetry parameters.
In general we find
\begin{equation}\label{BPS}
\delta {\cal{L}}_{{\rm brane}}\,\sim\, \bigl (
\tau_{\rm brane}\mathbbm{1} + f \gamma_{01\cdots p}\,\sigma\bigr ) \epsilon\,.
\end{equation}
This variation is proportional to a projection operator
provided
\begin{equation}\label{tension}
\tau_{\rm brane} =  f \,.
\end{equation}
We conclude
that requiring supersymmetry in this sector not only determines
the brane tension, via \eqref{tension}, but also fixes the BPS condition
on the supersymmetry parameter, via \eqref{BPS} and
\eqref{tension}. Note that, in order
to have a projection operator in \eqref{BPS} we must have
$\sigma = \sigma_1$ or $\sigma=\sigma_3$ for $p=1,5,9$ and
$\sigma=i\sigma_2$ for $p=3,7$.

Instead of terms linear in the gravitino, one can also consider
terms linear in the dilatino. This requires varying the brane tension in
front of the Nambu-Goto term. In all cases, except for nine-branes (see
section \ref{nine-branes}) we find that all dilatino
terms cancel provided that the same
projection operator and the same brane tension is used that follows from
requiring the cancellation of the gravitino terms. This provides a
non-trivial check on our calculations.

\section{Strings, Three-branes and Five-branes}

As a key example we consider the fundamental
string F1 and the Dirichlet brane D1. These branes form a doublet under
$SL(2,\mathbb{R})$ in which D1 is the S-dual of
F1\,\footnote{In this paper we mean by S-duality the discrete $\mathbb{Z}_2$ transformation.}.

In a formulation where the world volume vector field has been
integrated out\footnote{Integrating out the worldvolume vector field is
optional. However, to preserve $SL(2,\mathbb{R})$ symmetry one should
do this for both or none of the two branes.}
for both the F1 and
the D1 \cite{Townsend:1997kr},
the Nambu-Goto and Wess-Zumino terms are given by
($\phi$ is the dilaton and $\ell$ is the axion)\footnote{In this letter
all $p$-forms are real. With respect to \cite{Bergshoeff:2005ac}
in some cases the $p$-form is multiplied by a factor of  $i$. In the present
case $C=iC_{\rm old}$.}:
\bea
{\cal{L}}_{\rm D1} &=& \tau_{\rm D1} \sqrt{-g} +
\tfrac{1}{4}\epsilon^{\mu\nu} C_{\mu\nu}\,,\\
&&\nonumber\\
{\cal{L}}_{\rm F1} &=&  \tau_{\rm F1}\sqrt{-g} + \tfrac{1}{4}
\epsilon^{\mu\nu}
B_{\mu\nu}\,.
\eea
Here $C_{\mu\nu}$ and $B_{\mu\nu}$ are two-form potentials
that transform as a doublet under $SL(2,\mathbb{R})$.
The relevant supersymmetry rules of these potentials are given by
\be
\delta B_{\mu\nu} = 8i\bar\epsilon\,\sigma_3\gamma_{[\mu}\psi_{\nu]}\,,
\hskip 1truecm \delta C_{\mu\nu} = -8ie^{-\phi}\bar\epsilon\sigma_1
\gamma_{[\mu}\psi_{\nu]} +\ell\delta B_{\mu\nu}\,.
\ee
This can be used to determine the tensions and the BPS conditions of
the F1- and D1-branes, as explained above. In this way we find for
the fundamental string:
\be
\tau_{\rm F1} = 1\,,\hskip 2truecm \tfrac{1}{2}\bigl (\mathbbm{1} + \sigma_3
\gamma_{01}\bigr ) \epsilon = 0\,.
\ee
The analysis of the D1-brane is slightly more subtle because,
due to the $\ell\delta B_{\mu\nu}$ term in the variation of $C_{\mu\nu}$,
the variation of the D1 Wess-Zumino term contains both terms with a
$\sigma_1$ and a $\sigma_3$ matrix:
\be
\delta {\cal {L}}_{\rm D1} \sim \bigl (\tau_{\rm D1}\mathbbm{1} + (f\sigma_1
+ g\sigma_3)\gamma_{01}\bigr )\,
\epsilon\,,
\ee
with $f= -e^{-\phi}$ and $g= \ell$. Since $\sigma_1$ and
$\sigma_3$ anticommute we  find instead of \eqref{tension}:
\be \label{tension2}
\tau_{\rm D1} = \sqrt{f^2 + g^2} = \sqrt{e^{-2\phi} + \ell^2}\,.
\ee
Our results so far are summarized in table \ref{Table1}. The table
also contains the results for the 3- and 5-branes which can be
derived similarly.

\begin{table}[ht]
\begin{center}
\hspace{-1cm}
\begin{tabular}{|c||c|c|c|}
\hline \rule[-1mm]{0mm}{6mm}
potential & brane & tension &projection operator\\
\hline \rule[-1mm]{0mm}{6mm}
$ C_{(2)}$ & D1& $\sqrt{e^{-2\phi}+\ell^2}$&
$\frac{1}{2}\bigl(\mathbbm{1}+
\frac{-e^{-\phi} \s_1 + \ell \s_3}{\sqrt{e^{-2\phi}+ \ell^2}}\gamma_{01}\bigr)$\\
\hline \rule[-1mm]{0mm}{6mm}
$B_{(2)}$ & F1 & 1&$\tfrac{1}{2}\bigl(\mathbbm{1} + \sigma_3\gamma_{01}\bigr)$\\
\hline \rule[-1mm]{0mm}{6mm}
$C_{(4)}$ & D3  &   $e^{-\phi}$     &   $\tfrac{1}{2}(\mathbbm{1} + i \s_2\g_{0123})$\\
\hline \rule[-1mm]{0mm}{6mm}
$C_{(6)}$ & D5  &   $e^{-\phi}$     &   $\tfrac{1}{2}(\mathbbm{1} + \s_1 \g_{01 \cdots 5})$\\
\hline \rule[-1mm]{0mm}{6mm}
$B_{(6)}$ & NS5 &$e^{-\phi} \sqrt{e^{-2\phi}+ \ell^2}$  &
$\frac{1}{2}\bigl(\mathbbm{1}+ \frac{e^{-\phi} \s_3 + \ell\sigma_1
}{\sqrt{e^{-2\phi}+ \ell^2 }}\g_{01\cdots5}\bigr)$\\
\hline
\end{tabular}
\caption{The two 2-, 4- and 6-form potentials of IIB supergravity and the corresponding
branes in string frame. \label{Table1}}
\end{center}
\end{table}

We now consider a $(p,q)$-string (see \cite{Schwarz:1995dk}) where F1 denotes a $(1,0)$-string and
D1 a $(0,1)$-string:
\be
{\cal{L}}_{\rm (p,q)} = \tau_{(p,q)}\,\sqrt{-g} +
\tfrac{1}{4}\epsilon^{\mu\nu}\bigl (p\, B_{\mu\nu} + q\,C_{\mu\nu}\bigr )\,.
\ee
Again we find in the variation of ${\cal{L}}_{(p,q)}$ a combination
of two Pauli matrices:
\be
\delta {\cal{L}}_{(p,q)} \sim \bigl ( \tau_{(p,q)}\mathbbm{1} +
((p+\ell q) \sigma_3
- e^{-\phi}q\sigma_1)\gamma_{01}\bigr )\, \epsilon\,.
\ee
The tension now becomes
\be \label{taupq}
\tau_{p,q} = \sqrt{(p+\ell q)^2 + e^{-2\phi} q^2}\,,
\ee
which reproduces the tension formula of \cite{Schwarz:1995dk}.
Using Einstein frame this tension formula can be rewritten
in the manifest $SL(2,\mathbb{R})$-invariant form\footnote{Note that
$\tau_{\rm string} = e^{-\tfrac{p+1}{4}\phi}\tau_{\rm Einstein}$ for a
$p$-brane.}
\be \label{Einsteinstring}
\tau_{(p,q)}^{\rm E} = \sqrt{q^\alpha q^\beta {\cal M}_{\alpha\beta}}\,,
\ee
with
\be
q^\alpha = \begin{pmatrix}
q \cr p
\end{pmatrix}\hskip.5truecm {\rm and}\hskip .5truecm {\cal M} = e^\phi
\begin{pmatrix}
\ell^2 + e^{-2\phi} &\ell\cr
\ell & 1
\end{pmatrix}\,.
\ee

\section{Seven-branes}

Before considering nine-branes, it is instructive to first consider
seven-branes. There are three eight-form potentials\footnote{There is a constraint on the nine-form field strengths
(see \cite{Meessen:1998qm,Dall'Agata:1998va,Bergshoeff:2005ac}), such
that the three eight-forms describe the same two propagating degrees of freedom
as the dilaton and axion.}
$C_{(8)}, D_{(8)}$ and $B_{(8)}$
that transform as a triplet under $SL(2,\mathbb{R})$, see table \ref{Table2}.
They correspond to the three seven-branes D7, I7 and ${\widetilde {\rm D7}}$,
where ${\widetilde {\rm D7}}$ is minus the S-dual of D7 and
I7 changes sign under under S-duality.
Note that all seven-branes have the same projection operator and hence
correspond to the single 3-form (or, equivalently, 7-form)
central charge in the IIB supersymmetry
algebra. This shows that a single central charge that is
invariant under S-duality may correspond to different branes that
transform non-trivially under S-duality.

\begin{table}[ht]
\begin{center}
\hspace{-1cm}
\begin{tabular}{|c||c|c|c|}
\hline \rule[-1mm]{0mm}{6mm}
potential & brane & tension &projection operator\\
\hline \rule[-1mm]{0mm}{6mm}
 $C_{(8)}$ & D7& $e^{-\phi}$&$\tfrac{1}{2}\bigl (\mathbbm{1} + i\gamma_{01\cdots 7}
\sigma_2\bigr)$\\
\hline \rule[-1mm]{0mm}{6mm}
 $D_{(8)}$ & I7 & $\ell e^{-\phi}$&$\tfrac{1}{2}\bigl(\mathbbm{1} + i\gamma_{01\cdots 7}\sigma_2
\bigr)$\\
\hline \rule[-1mm]{0mm}{6mm}
$B_{(8)}$ & ${\widetilde {\rm D7}}$ &
$e^{-3\phi} + \ell^2e^{-\phi}$&$\tfrac{1}{2}\bigl (\mathbbm{1} + i\gamma_{01\cdots 7} \sigma_2\bigr )$\\
\hline
\end{tabular}
\caption{The three eight-form potentials of IIB supergravity and the
corresponding seven-branes. \label{Table2}}
\end{center}
\end{table}

Consider
now a combination of seven-branes:
\be
{\cal{L}}_{(p,r,q)} \sim \tau_{(p,r,q)}\, \sqrt{-g} +
\epsilon^{\mu_1\cdots \mu_8}\bigl( p \, C_{\mu_1\cdots \mu_8}
+ r D_{\mu_1\cdots \mu_8} + q B_{\mu_1\cdots \mu_8}\bigr)\,.
\ee
The D7-brane corresponds to $(p,r,q) = (1,0,0)$, the others accordingly.
We find that in the supersymmetry variation the terms linear in the
gravitino are proportional to
\be
\delta {\cal{L}}_{(p,r,q)} \sim \bigl (\tau_{(p,r,q)} \mathbbm{1} +
i\bigl (p\,e^{-\phi} + r\, \ell e^{-\phi} + q\, (e^{-3\phi} + \ell^2e^{-\phi})
\bigr )\gamma_{01\cdots 7}\sigma_2\bigr)\epsilon\,.
\ee
This is proportional to a projection operator provided that
\be
\tau_{(p,r,q)} =
|p\,e^{-\phi} + r\, \ell e^{-\phi} + q\, (e^{-3\phi} + \ell^2e^{-\phi})|\,.
\ee
Using Einstein frame this formula can be written in
manifest $SL(2,\mathbb{R})$-invariant form as follows:
\be
\tau_{(p,r,q)}^{\rm E} = |q^{\alpha\beta} {\cal{M}}_{\alpha\beta}|\,,
\ee
with
\be
q^{\alpha\beta} = \begin{pmatrix}
q&r/2\cr
r/2&p
\end{pmatrix}\,.
\ee
In contrast to strings we can impose an $SL(2,\mathbb{R})$-invariant
constraint on the matrix $q^{\alpha\beta}$:
\be \label{constraint}
{\rm det}\, [q^{\alpha\beta}]\ = -\alpha^2\hskip .5truecm {\rm or}
\hskip .5truecm  pq - \frac{r^2}{4} = - \alpha^2\,,
\ee
for some $\alpha$.
We see that the D7-brane and the $\widetilde{\rm D7}$-brane
belong to the $\alpha=0$ conjugacy class but that I7 belongs to the
$\alpha^2>0$ conjugacy classes.
The constraint \eqref{constraint} defines co-dimension 1 surfaces
in $\mathbb{R}^{2,1}$. For $\alpha \ne 0$ they are hyperboloids and for
$\alpha=0$ a cone. These hypersurfaces are homogeneous spaces\footnote{We thank
Patrick Meessen for discussions of this point.}, and, therefore,
all of them can be constructed as coset spaces $SL(2,\mathbb{R})/H_\alpha$
where $H_\alpha$ is the isotropy group for a given $\alpha$.
For instance, for $\alpha=0$, we find that  $H_0$
is the subgroup of matrices of the form
\be \label{isotropy}
\Lambda = \begin{pmatrix}
1&&b\cr 0&& 1
\end{pmatrix}\ee
that shift the value of $\ell$ by a real constant, which is isomorphic to
$\mathbb{R}$.

We can use the constraint \eqref{constraint}
to solve for $r$ in terms of $p$ and $q$:
\be
r(p,q) = \pm 2\sqrt{pq+\alpha^2}\,.
\ee
This provides us with a set of $(p,q)$ seven-branes that define a
two-dimensional ma\-nifold. For $\alpha^2=0$ this manifold is parametrized
by
\be
(p,r,q) = (b^2, 2bd, d^2)\,,\hskip 1.5truecm b,d \in \mathbbm{R}\,.
\ee

The $\alpha^2>0$ and $\alpha^2<0$ $(p,q)$ seven-branes form
distinct conjugacy classes whose elements cannot be obtained by any
$SL(2,\mathbb{R})$ transformation of the D7-brane which belongs to the
$\alpha^2=0$ conjugacy classes. Nevertheless, for each $\alpha$
they represent supersymmetric seven-branes whose solutions
have been constructed \cite{Bergshoeff:2002mb}. Using a basis with $r=0$ and
restricting ourselves to  $\alpha^2 = 0, \pm 1$ we can
write a representative for each conjugacy class as follows:
\be
\alpha^2=0\,: \begin{pmatrix}1\cr 0\cr 0\cr\end{pmatrix}\,,\hskip 1truecm
\alpha^2=-1\,: \begin{pmatrix}1\cr 0 \cr 1\cr\end{pmatrix}\,,\hskip 1truecm
\alpha^2=1\,: \begin{pmatrix}1\cr 0\cr -1\cr\end{pmatrix}\,,
\ee
where the 3-vector indicates a vector with components $(p,r,q)$.
This shows that elements of the $\alpha^2 = 1$ and  $\alpha^2 = -1$ conjugacy
class correspond
to duality transformations of bound states of D7 branes with
$\widetilde {\rm D7}$ branes and $\widetilde {\overline {\rm D7}}$ branes
(anti S-dual D7-branes), respectively.

\section{Nine-branes}\label{nine-branes}

We now consider the main topic of this letter: nine-branes.
As explained in \cite{Bergshoeff:2005ac} the ten-form potentials of
IIB supergravity organize themselves in a doublet and
quadruplet representation of $SL(2,\mathbb{R})$. Note that the
10-form potential $C_{(10)}$ that couples to the D9-brane is in the
quadruplet representation. Using the supersymmetry
rules given in \cite{Bergshoeff:2005ac} it is straightforward to
determine the tensions and BPS conditions of the different nine-branes
by requiring the cancellation
of all terms linear in the gravitino in the supersymmetry variation of the
different 9-brane actions. The results for the doublet and quadruplet are
summarized in tables \ref{doublet} and \ref{quadruplet}, respectively.
\bigskip

\begin{table}[ht]
\begin{center}
\begin{tabular}{|c||c|c|c|}
\hline \rule[-1mm]{0mm}{6mm}
potential &brane & tension & projection operator\\
\hline \rule[-1mm]{0mm}{6mm}
${\cal{D}}_{(10)}$ &S9  &   $e^{-2\phi}$
    &$\tfrac{1}{2}(\mathbbm{1} + \s_3)$\\
\hline \rule[-1mm]{0mm}{6mm}
${\cal{E}}_{(10)}$ & $\widetilde{\rm S9}$   &   $e^{-2\phi}
\sqrt{e^{-2\phi} + \ell^2}$ & $\tfrac{1}{2}\bigl(
\mathbbm{1}+ \tfrac{- e^{-\phi} \s_1 + \ell
\s_3}{\sqrt{e^{-2\phi} + \ell^2}}\bigr)$\\
\hline
\end{tabular}
\caption{The doublet of 10-form potentials, their tensions and their
projection operators. \label{doublet} }
\end{center}
\end{table}

\begin{table}[ht]
\begin{center}
\begin{tabular}{|c||c|c|}
\hline \rule[-1mm]{0mm}{6mm}
potential &brane & tension $\t$ and projection operator $P$\\
\hline \rule[-1mm]{0mm}{6mm}
\multirow{2}{*}{$C_{(10)}$}&\multirow{2}{*} {D9}    &   $\t = e^{-\phi}$ \\
& &$P=\tfrac{1}{2}(\mathbbm{1} + \s_1)$\\
\hline \rule[-1mm]{0mm}{6mm}
\multirow{2}{*}{$D_{(10)}$}&\multirow{2}{*} {--} &  $
\t=e^{-\phi}\sqrt{\tfrac{1}{9}\,e^{-2\phi}+  \ell^2}$\\
& & $P=\tfrac{1}{2} \bigg(\mathbbm{1} + \frac{\ell\,\s_1 + \tfrac{1}{3}\,e^{-\phi}
\s_3}{\sqrt{\tfrac{1}{9}\,e^{-2\phi}+  \ell^2 }}\bigg)$\\
\hline \rule[-1mm]{0mm}{6mm}
\multirow{2}{*}{$E_{(10)}$}& \multirow{2}{*}{--} &  $
\t=e^{-\phi}\sqrt{(\tfrac{1}{3}e^{-2\phi} + \ell^2 )^2+
\tfrac{4}{9}\, \ell^2
e^{-2\phi}}$ \\ & & $P=\tfrac{1}{2} \bigg(\mathbbm{1}-
 \frac{( \tfrac{1}{3}\,e^{-2\phi} +
\ell^2  )\s_1 +
\tfrac{2}{3}\, \ell e^{-\phi} \s_3}{\sqrt{(\tfrac{1}{3}\,e^{-2\phi}+
\ell^2 )^2+ \tfrac{4}{9}\,\ell^2 e^{-2\phi}}}\bigg)$\\
\hline \rule[-1mm]{0mm}{6mm}
\multirow{2}{*}{$B_{(10)}$}&\multirow{2}{*}{${\widetilde{\rm D9}}$} &
$\t=e^{-\phi} \bigl(e^{-2\phi} +\ell^2
\bigr)^{3/2}$  \\ & & $P=\frac{1}{2} \bigl( \mathbbm{1}-
\frac{ \ell \s_1 +
e^{-\phi} \s_3}{\sqrt{e^{-2\phi} + \ell^2}}\bigl)$\\
\hline
\end{tabular}
\caption{The quadruplet of 10-form potentials, their tensions and their
projection operators. Note that there is no half-supersymmetric
nine-brane that couples to $D_{(10)}$ or $E_{(10)}$.
\label{quadruplet} }
\end{center}
\end{table}

The discussion of the doublet of nine-branes is rather similar to that
of the doublet of strings. We find that the tension of a $(p,q)$
nine-brane is given by
\be
\tau_{(p,q)} = e^{-2\phi} \sqrt{(p+\ell q)^2 + e^{-2\phi} q^2}\,.
\ee
In Einstein frame the tension is given by the manifest
$SL(2,\mathbb{R})$-invariant tension-formula \eqref{Einsteinstring}.

The discussion of the quadruplet of nine-branes is more subtle.
We first consider $(p,r,s,q)$-branes
\begin{equation}
{\cal{L}}_{(p,r,s,q)} \sim \tau_{(p,r,s,q)}\, \sqrt{-g} +
\epsilon^{\mu_1\cdots \mu_{10}}\bigl( pC_{\mu_1\cdots \mu_{10}}
+ r D_{\mu_1\cdots \mu_{10}} + s E_{\mu_1\cdots \mu_{10}}+
 q B_{\mu_1\cdots \mu_{10}}\bigr)\,,
\end{equation}in which the $(1,0,0,0)$-brane
represents the D9-brane etc. We find
that the tension of a $(p,r,s,q)$-brane is given by
\bea
\tau_{(p,r,s,q)} &=& \left[ \{ e^{-\phi}p+\ell e^{-\phi}r-
\left(\tfrac{1}{3}e^{-3\phi}+\ell^2e^{-\phi}\right)s-
\left(\ell^3 e^{-\phi} + \ell e^{-3\phi}\right)q \}^2\nonumber \right.\\
&&+ \left. \{ \tfrac{1}{3}e^{-2\phi}r- \tfrac{2}{3}\ell e^{-2\phi}s- \left(e^{-4\phi} +
\ell^2 e^{-2\phi}\right)q \}^{2} \right]^{1/2}\,.
\eea
In Einstein frame the manifest $SL(2,\mathbb{R})$-invariant
tension is given by
\be
\tau_{(p,r,s,q)}^{\rm E} = \sqrt{q^{\alpha\beta\gamma} q^{\delta\epsilon\zeta}
{\cal{M}}_{\alpha\beta}{\cal{M}}_{\delta\epsilon}{\cal{M}}_{\gamma\zeta}}\,,
\ee
where
\be
q^{222} = p\,,\hskip .5truecm
q^{122} = -\tfrac{1}{3}r\,,\hskip .5truecm
q^{112} = -\tfrac{1}{3}s\,,\hskip .5truecm
q^{111} = q\,.
\ee

So far we have only achieved the cancellation of the gravitino terms.
These cancellations merely serve to derive the tension formulae and
the BPS conditions. A first nontrivial check on supersymmetry is the
cancellation of the dilatino terms. Unlike the previous cases
we find that these dilatino terms only cancel
provided that
\be \label{2constraints}
3qr+s^2=0\,,\hskip 1truecm 3ps+r^2=0\,, \hskip 1truecm
9pq -rs =0\,.
\ee
Note that these constraints are  satisfied by the
D9-brane and its S-dual but {\it not} by the other two 9-branes.
For generic points $(p,r,s,q)$ the last constraint follows from the
first two but not for special cases. Therefore, it cannot
be omitted. For instance, the points $(p,0,0,q)$
solve the first two constraints for general $p,q$ but solving
the third constraint requires $p=0$ or $q=0$.

To understand the $SL(2,\mathbb{R})$ properties of the constraints
\eqref{2constraints} it is convenient to introduce the matrix
$
Q^{\alpha\beta} \equiv q^{\alpha\gamma\delta}q^{\beta\epsilon\zeta}
\epsilon_{\gamma\epsilon}\epsilon_{\delta\zeta}$ or
\be
Q^{\alpha\beta} = \frac{1}{9}\begin{pmatrix}
2(3qr+s^2)&9pq-rs\cr
9pq-rs&2(3ps+r^2)
\end{pmatrix}\, .
\ee
The constraints \eqref{2constraints} can then be written as
$Q^{\alpha\beta} = 0$ and transform manifestly as a triplet of $SL(2,\mathbb{R})$.
Using the constraints \eqref{2constraints} to solve for $r,s$ in terms
of $p,q$ we end up with a set of $(p,q)$ 9-branes
that define a two-dimensional
manifold in a four-dimensional space parametrized by
\be
(p,r,s,q) = (d^3, -3bd^2, -3b^2 d, b^3)\,,\hskip 1,5truecm c,d
\in \mathbb{R}\,. \label{D9orbit}
\ee
This manifold
can be identified as the $SL(2,\mathbb{R})$ orbit of the D9-brane.
It has the isotropy group \eqref{isotropy}. We thus end up
with the same homogeneous space that we encountered for the
orbit of the D7-brane. Unlike the case of seven-branes there are no
other conjugacy classes of half-supersymmetric nine-branes.

Finally, we consider the "quantized"
duality group $SL(2,\mathbb{Z})$. We assume that the brane
charges are quantized. We first consider the linear doublet.
The orbit of the S9-brane is given by:
\be
\begin{pmatrix}
a&&b\\
c&&d
\end{pmatrix}
\begin{pmatrix}1\\0\end{pmatrix} = \begin{pmatrix}a\\c\end{pmatrix}
\,\, , \,\,\, ad -bc =1. \ee
For any pair $a,c$ of co-prime integers\footnote{If $a,c$ are not
 co-prime we consider
them as multiple copies of a fundamental brane, see \cite{Schwarz:1995dk}.}
we can use the
extended Euclidean algorithm to solve for integers $b$ and $d$. This
shows that if we
restrict the duality group to
$SL(2,\mathbb{Z})$, all branes are  in the orbit of the
S9-brane.
Note that the same argument applies to $(p,q)$-strings and $(p,q)$
5-branes. The
case of $(p,q)$ 7-branes, where different classes of constraints appear,
is more complex and was treated in \cite{DeWolfe:1998pr}.

We next consider the non-linear doublet, where we find a similar result:
 all 9-branes lie in the
$SL(2,\mathbb{Z})$ orbit of a single D9-brane, if we do not count multiple
branes of the same type as independent. This can be seen as follows\footnote{A similar
analysis, with a similar result, can be made of the
$\alpha^2=0$ conjugacy class
of seven-branes, which contains the D7-brane.}
. All
9-branes lie in the $SL(2,\mathbb{R})$ orbit of
the D9-brane, given by the charges $(d^3, -3bd^2, -3b^2 d, b^3)$.
To verify our
claim we have to check
that all $SL(2,\mathbb{R})$ transformations of the D9-brane
which lead to integer charges are
also elements of $SL(2,\mathbb{Z})$. To obtain integer charges, we
must have $d^3 = u, \,\, d^2b = v$ for some integers $u,v$. If $d$ and $b$ are not
integers themselves, then without loss of generality this implies
$b = n d$ for some integer $n$\,\footnote{If only one of them is not an integer, then we do
not get integer charges and if both are integers, then there is either an
$SL(2,\mathbb{Z})$ transformation which generates this brane from the D9-brane, or
we are dealing with a multiple brane.}. Any transformation of that kind, however, leads to a "multiple" brane
with charges $u (1,-3 n,-3 n^2,n^3)$, which we do not consider independent,
unless $u=d=1,\,\,b=n$. In the latter case the brane is connected to the D9-brane
by a
$SL(2,\mathbb{Z})$  transformation\footnote{To construct this
transformation, one can determine the values of $a,c$ with
the extended Euclidean algorithm, similarly to
the S9-brane case.}. This verifies our claim.

\section{Discussion}

In this work we have determined the tensions of all
half-supersymmetric branes of IIB string theory,
including a linear and nonlinear
doublet of $(p,q)$ nine-branes. A brief summary of the results is given in
table \ref{AllTensions}.

\begin{table}[ht]
\begin{center}
\begin{tabular}{|c||c|c|c|c|c|c|c|c|c|c|c|c|c|}
\hline \rule[-1mm]{0mm}{6mm}
brane   & $\rm D1$          & $\rm F1$& $\rm D3$        & $\rm D5$      & $\rm NS5$     & $\rm D7$              & $\widetilde {\rm D7}$ & $\rm D9$ &
 $\widetilde{\rm D9}$& S9 &  $\widetilde{\rm S9}$\\
\hline \rule[-1mm]{0mm}{6mm}
tension & $g^{-1}_s$    & $1$ & $g^{-1}_s$  & $g^{-1}_s$& $g^{-2}_s$& $g^{-1}_s$&  $g^{-3}_s$ & $g^{-1}_s$ & $g^{-4}_s$ & $g^{-2}_s$ & $g^{-3}_s$\\
\hline
\end{tabular}
\caption{The tensions of the basic  half-supersymmetric IIB branes,
with vanishing axion. In the case of seven-branes we have not indicated the
half-supersymmetric branes corresponding to the $\alpha^2>0$ and $\alpha^2<0$
conjugacy classes of $SL(2,\mathbb{R})$.
\label{AllTensions} }
\end{center}
\end{table}

There are some surprises for the nine-branes. The standard D9-brane
is part of a nonlinear doublet. This nonlinear doublet is expected
to play a role in the construction  of strings with sixteen
supercharges, along the lines of Refs.
\cite{hull1,hull2,9812224}\footnote{Observe, however, that in these
papers the D9-brane was assumed to belong to a linear doublet
together with the $\widetilde{\rm D9}$ (there named NS9), which was
shown to have a tension proportional to $g_s^{-4}$ in the string
frame. We have shown here that this doublet is non-linear, with the
corresponding potentials belonging to a quadruplet.}. We find an
additional linear doublet of nine-branes that contains a S9-brane
whose tension scales as $g_s^{-2}$, i.e.~a soliton. An obvious
question to ask is: what is the world-volume dynamics of the
S9-brane? We hope to report in this direction in a forthcoming paper
\cite{upcoming}. The precise role of the S9-brane in IIB string
theory is still unclear. Perhaps it has a role to play in the
recently proposed non-perturbative open SO(32) heterotic string
construction of \cite{Polchinski:2005bg}\, \footnote{We thank Timo
Weigand for pointing out this reference to us.}.

\section*{Acknowledgements}

 We wish to thank Diederik Roest for an interesting discussion on
brane tensions and Ana Mar\'{\i}a Font for a remark about the S-duality
of 7-branes. E.B., M. de R., S.K.~and T.O.~are supported
by the European Commission FP6 program
MRTN-CT-2004-005104 in which E.B., S.K. and M. de R. are
associated to Utrecht university.
S.K. is supported by a Postdoc-fellowship of the German
Academic Exchange Service (DAAD).
F.R. is supported by a European
Commission Marie Curie Postdoctoral Fellowship, Contract
MEIF-CT-2003-500308 and thanks the INFN for support and the Department of
Physics for hospitality at University of Rome `Tor Vergata'.
The work of E.B.~and T.O.~is supported by the Spanish grant
BFM2003-01090.

\end{document}